\documentclass[preprint,12pt,authoryear]{elsarticle}

\usepackage{amssymb}
\usepackage{amsmath}
\usepackage{multirow}
\usepackage{booktabs}
\usepackage[figuresright]{rotating}
\usepackage{tabularx}
\usepackage{float}

\journal{International Journal of Human-Computer Interaction}

\begin{document}

\begin{frontmatter}



\title{Enhancing Psychotherapeutic Alliance in College: When and How to Integrate Multimodal Large Language Models in Psychotherapy} 


\author[label1]{Jiyao Wang}
\author[label1]{Youyu Sheng}
\author[label2]{Qihang He}
\author[label1]{Haolong Hu}
\author[label3]{Shuwen Liu}
\author[label1]{Feiqi Gu}
\author[label4]{Yumei Jing\corref{cor1}}
\author[label1]{Dengbo He\corref{cor1}}
\affiliation[label1]{organization={The Hong Kong University of Science and Technology (Guangzhou)},
            city={Guangzhou},
            state={Guangdong},
            country={China}}
\affiliation[label2]{organization={Sichuan University},
            city={Chengdu},
            state={Sichuan},
            country={China}}
\affiliation[label3]{organization={Nanjing University of Science and Technology},
            city={Nanjing},
            state={Jiangsu},
            country={China}}
\affiliation[label4]{organization={Hubei Normal University},
            city={Huangshi},
            state={Hubei},
            country={China}}
\cortext[cor1]{Corresponding author, email address: jingyumei@hbnu.edu.cn; dengbohe@hkust-gz.edu.cn}
\begin{abstract}
As mental health issues rise among college students, there is an increasing interest and demand in leveraging Multimodal Language Models (MLLM) to enhance mental support services, yet integrating them into psychotherapy remains theoretical or non-user-centered. This study investigated the opportunities and challenges of using MLLMs within the campus psychotherapy alliance in China. Through three studies involving both therapists and student clients, we argue that the ideal role for MLLMs at this stage is as an auxiliary tool to human therapists. Users widely expect features such as triage matching and real-time emotion recognition. At the same time, for independent therapy by MLLM, concerns about capabilities and privacy ethics remain prominent, despite high demands for personalized avatars and non-verbal communication. Our findings further indicate that users' sense of social identity and perceived relative status of MLLMs significantly influence their acceptance. This study provides insights for future intelligent campus mental healthcare.
\end{abstract}



\end{frontmatter}



\section{Introduction}
Mental disorders remain among the top 10 causes of burden globally, with no evidence of a reduction globally since 1990 \citep{rudd2020global}. In 2019, the World Health Organization (WHO) reported that approximately 970 million people suffered from mental health disorders \citep{who1}. Particularly, as late adolescence or early adulthood (e.g., alcohol and substance abuse \citep{weitzman2004poor} are more sensitive to emotional stimuli \citep{iarovici2014mental, balon2015college}), college students have an even higher incidence of mental disorders \citep{emmons2014fall,hunt2010mental,kessler2005lifetime}. For example, in the U.S., depression affects 17.3\% of undergraduates, and anxiety and depression often co-occur \citep{eisenberg2013mental,liu2019prevalence}. In response to students' increasing mental health concerns, traditionally, face-to-face campus counseling and mental health services are considered an effective way \citep{zabek2023roles}. However, many college students fail to seek timely help even when counseling centers are available on campus \citep{hunt2010mental}. The reasons include social stigma \citep{barney2006stigma,baumeister2012inappropriate}, personal emotional barriers, and financial challenges associated with face-to-face counseling \citep{dieleman2016us}. Further, in some special periods, for example during the COVID-19 pandemic, physical isolation hinders college students from seeking and receiving help \citep{ebert2019barriers,carlbring2018internet}.

With the rapid development of artificial intelligence (AI), researchers attempted to integrate AI into psychotherapy to support those in need. Particularly, as an effective information extraction tool, AI can also be used to accelerate the identification and treatment of mental disorders for both psychotherapists and users \citep{AKTAN2022107273,liu2022using}. For instance, text sentiment analysis can understand users' current state quicker, which can assist psychotherapists in making more accurate interventions, and allows users to identify their emotional problems earlier and seek mental health support in a timely manner \citep{jeong2023deploying}. Moreover, the use of chatbots driven by natural language processing (NLP) and machine learning (ML) complement the role of clinicians \citep{haque2018measuring}. These chatbots mimic human conversations, allowing individuals to interact with them for support and guidance on their mental health needs at any time \citep{ware2020predicting, bowman2024exploring}, and anywhere (e.g., at home via users' own mobile devices \citep{mastoras2019touchscreen}). Additionally, being different from human counselor-patient therapy \citep{apa}, some researchers \citep{liu2022using, lee2024influence,jeong2023deploying} have also attempted to construct conversational AI agents that can provide autonomous therapy. This kind of agent is expected to have the ability to independently deliver complete psychotherapeutic services like a human expert. Despite these, limited by the technology, conversational chatbots or AI tools based on traditional NLP is difficult to perceive the client's shifts of engagement and emotion from information in other modalities; therefore the therapeutic effect would be significantly reduced if no human was involved \citep{khanna2022affective}. 

Recently, the superior and impressive performance of the multi-modal large language model (MLLM) (e.g., GPT4-o\footnote{https://openai.com/index/hello-gpt-4o/}) in content generation \citep{wang2024evaluating, wang2024young}, sentiment analysis \citep{yang2024advancing}, and multi-modal perception \citep{huang2024language} proved their great potential in mental health compared to tools or chatbots powered by traditional NLP \citep{ai2023information}. Especially, based on the extremely high similarity obtained from a large number of studies comparing fine-tuned LLM-generated responses with human experts, researchers have begun to experiment with training LLM psychotherapists based on Cognitive behavioral therapy (CBT) \citep{nie2024llm,chiu2024computational}. 

Although the latest MLLMs with improved capabilities bring new opportunities, there are still concerns when they are used in the field of mental health. As concluded by \citep{grodniewicz2023waiting}, there are still no mature clinically integrated solutions for AI-enabled psychotherapy. Although there have been some attempts to evaluate existing integration schemes of MLLMs, most of them have been limited to analyzing them in terms of technical-type metrics (e.g., coherence of model-generated content, similarity to human experts, etc.) or theoretical discussions \citep{hua2024large,adhikary2024exploring,stade2024large}. However, psychotherapy always involves human interaction \citep{grodniewicz2023waiting}. Given that the introduction of MLLM will have a huge impact on the existing therapeutic alliance, the high prevalence of mental health problems in college students, and the greater acceptance of new technologies by young people \citep{broady2010comparison}, a comprehensive understanding of the attitudes, needs, and concerns of the stakeholders (i.e., human therapists and patients) regarding the adoption of AI in psychotherapy is critical to designing psychotherapy tools. Such understanding can speed up the validation and dissemination of new MLLM-based psychotherapy and thus benefit college mental health. 

In this study, we aim to answer the following research questions: (1) What are therapists' and patients' expectations of how to integrate current MLLMs into the psychotherapy workflow? (2) What functions should mental health MLLMs have to reach therapists' and patients' expectations? (3) What concerns do therapists and patients have about applying MLLMs to psychotherapy? To answer the above questions, we conducted focus group interviews with 15 therapists specializing in college student mental health, and semi-structured interviews with 20 college students who suffered from varying degrees of mental health problems. Besides, a larger-scale survey-based investigation was performed to supplement the broader attitudes of college students who had mental health problems. Together, in-depth discussions of the interview and survey data offered a nuanced perspective on opportunities, expectations, and challenges of integrating or using MLLMs for psychotherapy on college students. Our findings can provide insights into the function and application design of future MLLM-based mental health products for campus psychotherapy.

\section{Related Works}

\subsection{AI-Assisted Counseling Psychotherapy}
In the field of mental health, face-to-face campus counseling and mental health services are considered effective solutions to support students' increasing needs for mental health services \citep{zabek2023roles}. With the rise of AI, researchers have explored the adaptation of AI in psychotherapy activities. For example, machine learning techniques have been used in mental health tools and the research covered a variety of topics, including counselor adaptability and effectiveness \citep{perez2019makes}, the distinction between personalized and standardized counseling language \citep{althoff2016large}, shifts in psychological perspectives and topic adherence \citep{althoff2016large,wadden2021effect}, therapeutic interventions \citep{lee2019identifying}, empathy expression \citep{sharma2020computational}, transformative moments \citep{pruksachatkun2019moments}, counseling approaches \citep{perez2022pair,shah2022modeling}, and conversational involvement \citep{sharma2020engagement}. AI systems can analyze various types of behavioral data such as video, audio, and text from therapy sessions. This allows clinicians to receive immediate insight and feedback without relying on patients to complete extra self-report measures. Studies have shown that AI systems can accurately predict patient-reported alliance and symptom ratings based on behavioral markers extracted from video recordings of psychotherapy sessions. Verbal features \citep{aafjes2020language,vail2022toward,goldberg2020machine}, body and facial movements \citep{cohen2021nonverbal}, have all been considered as directly associated with self-reported measures of the working alliance.

Recently, impressed by the outstanding performance of LLM in many fields \citep{wang2024evaluating,kung2023performance,petridis2023anglekindling}, some studies tried to develop LLM-based tools to participate in traditional face-to-face psychotherapy. Particularly, in addition to classic functions like emotion recognition \citep{muller2024recognizing} and information retrieval \citep{stade2024large}, other advanced functions (e.g., LLM-based tools for counseling psychotherapists supervision \citep{li2024automatic}, conversational topic recommendation in counseling \citep{gunal2024conversational}, psychotherapists training \citep{wang2024patient}) are emerging.

However, although a small number of studies \citep{grodniewicz2023waiting, haber2024artificial} have mentioned the challenges of integrating large language models into traditional counseling modes, to the best of our knowledge, there is still a lack of user-centered (human psychotherapists and clients) research evaluating the current LLMs from the perspective of treating LLMs as supportive artificial intelligence systems in the context of human-computer interaction.

\subsection{Autonomous Psychotherapy by Conversational AI}

Given the advantages like ease of accessibility, and personalized service \citep{habicht2024closing}, researchers have also delved into developing virtual assistants and chatbots for autonomous health support \citep{richards2023principlist,nicol2022chatbot} and counseling services \citep{srivastava2023response}. Functionally, such AI systems can also be broadly classified into two categories. The first category provides daily counseling or chat services, including listening, empathy and reassurance, simulating the role of a friend or experienced elder \citep{xu2022survey,valtolina2021design}. For example, some studies \citep{harilal2020caro, kornfield2022meeting} tried to propose a digital tool for people with self-managing mental health concerns, and SERMO for emotion regulation \citep{denecke2020mental}. Meyerhoff et. al \citep{meyerhoff2022meeting} studied changes in the behavior of young people with depression who sought help from the community online, and provided recommendations for the corresponding design of future digital mental health tools. Other business products include OneRemission\footnote{https://keenethics.com/project-one-remission}, and Woebot\footnote{https://woebothealth.com/}. However, this type of chatbot still cannot replace professional psychotherapy services.

Another type of chatbot aims to mimic human therapists to replace them in existing counseling psychotherapy. However, numerous studies have expressed skepticism about the capability of this type of chatbot \citep{xu2022survey,brown2021ai}. Particularly, before 2023, most mental health chatbots were based only on traditional NLP techniques \citep{laranjo2018conversational}. The lack of visual information may lead to lower user engagement and a poor understanding of the user's current state due to the absence of facial emotion and movement analysis \citep{miner2019key}. 

With the development of AI, the incorporation of cognitive mechanisms has significantly enhanced the capabilities of LLMs \citep{binz2023turning}, which demonstrate emergent behaviors reminiscent of complex physical systems \citep{wei2022emergent}. Currently, based on the rapid development of LLM technology and its surprising capabilities, more and more works \citep{nie2024llm,chiu2024computational} have begun to target the second type of autonomous psychotherapeutic agent. However, despite the fast development of LLM after 2023, concerns about the capabilities of LLM-based chatbots \citep{grodniewicz2023waiting,floridi2023ai} still existed, though some have claimed that properly prompted LLMs are capable of complex reasoning \citep{wei2022chain} and even exhibit capabilities similar to psychologists \citep{ullman2023large}. One significant research gap for the adoption of the LLM into psychotherapy is that, rapid technology iteration does not necessarily lead to wide adoption, and the views of human participants as stakeholders in the different types of LLM-based autonomous psychotherapeutic agents are still missing from the extant literature in terms of more extensive discussions and analyses.


\section{Methodology}

\begin{figure*}
\begin{center}
\includegraphics[scale=0.5]{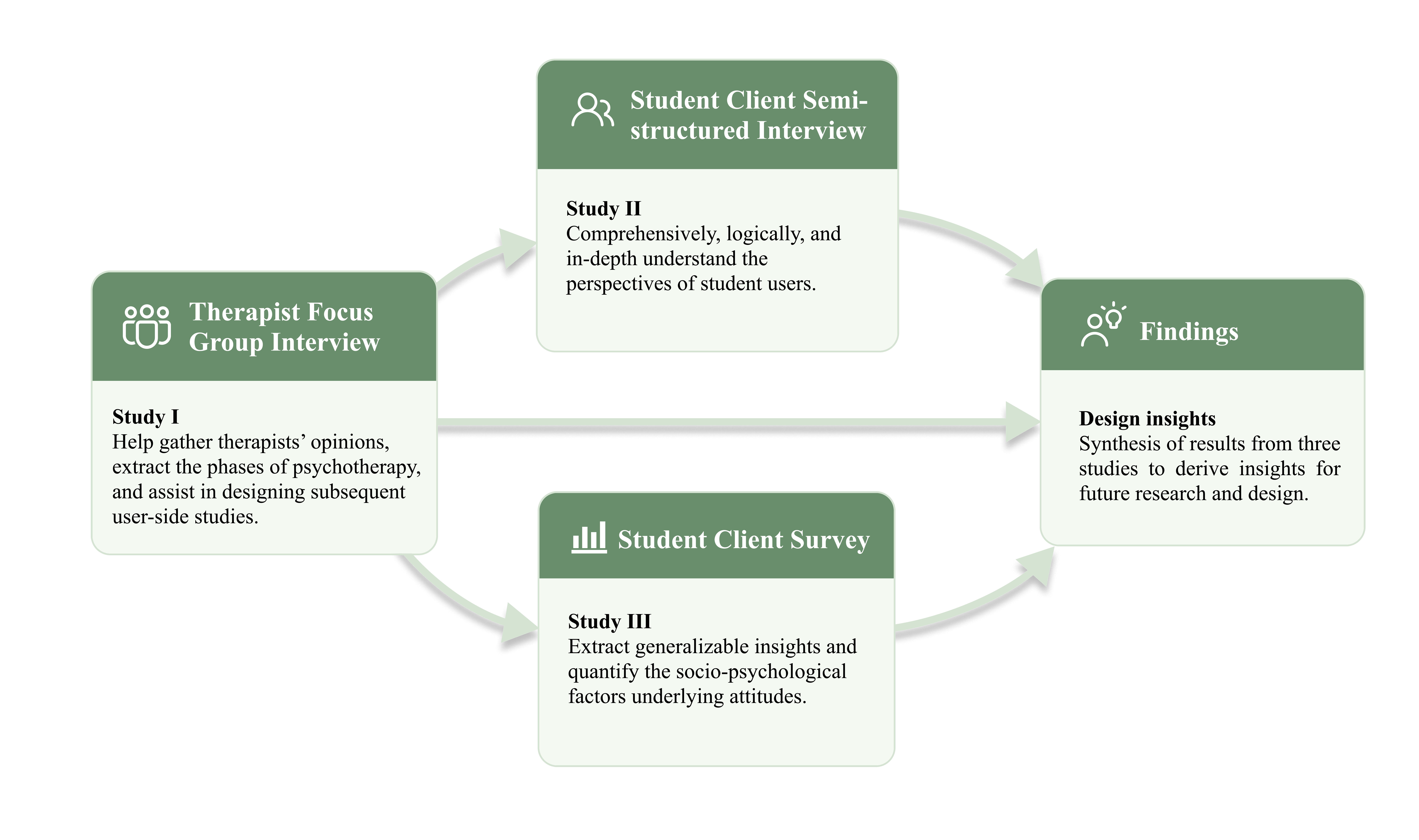}\\
\end{center}
\vspace{-4mm}
\caption{The flowchart of the study design.}\label{f0}
\vspace{-4mm}
\end{figure*}

In this study, we aimed to investigate proper ways of integrating MLLM into psychotherapy from the perspectives of users (i.e., psychotherapists and patients in college), and their demands and perceived barriers to applying MLLM for psychotherapy. Therefore, a mixed approach was adopted, consisting of a focus group interview on psychotherapists (Study I), a semi-structured interview with college students who had mental issues (Study II), and an online survey delivered to a wider range of college students (Study III). Specifically, the focus group and semi-structured interview were to gather human users' comments and suggestions about current MLLM-based solutions for mental health.  Then, considering a larger population of patients than psychotherapists, we collected opinions from college students through a questionnaire study. A visualization of our study design is shown in Figure \ref{f0}. The whole experiment was approved by the Ethics Committee of the Hubei Normal University.

\subsection{Participants}

\begin{table}[h]\centering
\caption{Demographic Information of the Interviewed Psychotherapists.}
\label{t1}
\scriptsize
\begin{tabular}{c|c|c|c|c}
\toprule
\textbf{ID} & \textbf{Age}& \textbf{Gender} & \textbf{Work years}& \textbf{Therapeutic Orientation}   \\
\hline
J1 & 28  & Female & 3-5             & Humanistic Therapy                                       \\
J2 & 27  & Female & 3-5             & Cognitive Behavioral Therapy                             \\
J3 & 28  & Female & 3-5             & Cognitive Behavioral Therapy                             \\
J4 & 29  & Female & 3-5             & Family Systems Therapy                                   \\
J5 & 27  & Male   & 3-5             & Humanistic Therapy                             \\
\hline
M1 & 34  & Female & 5-10            & Humanistic Therapy                                       \\
M2 & 38  & Female & 5-10            & Psychoanalytic Therapy, Humanistic Therapy               \\
M3 & 34  & Female & 5-10            & Humanistic Therapy, Solution-Focused Therapy             \\
M4 & 37  & Male   & 5-10            & Humanistic Therapy, Psychodynamic Therapy                \\
M5 & 31  & Female & 5-10            & Cognitive Behavioral Therapy                             \\
\hline
S1 & 60  & Female & $>$10              & Humanistic Therapy                                       \\
S2 & 51  & Male   & $>$10              & Psychodynamic Therapy                                    \\
S3 & 44  & Female & $>$10              & Humanistic Therapy, Family Systems Therapy               \\
S4 & 37  & Male   & $>$10              & Psychoanalytic Therapy, Cognitive Behavioral Therapy     \\
S5 & 47  & Female & $>$10              & Cognitive Behavioral Therapy, Mindfulness-Based Therapy       \\
\bottomrule

\end{tabular}
\textnormal{\\Notes: 'J' means the junior therapist group; 'M' stands for mid-career therapist; and 'S' is senior therapist group.}
\end{table}

\subsubsection{Study I: Focus Group Interview with Psychotherapists}
We recruited 15 psychotherapists (4 males and 11 females) who are currently working for different universities in mainland China and have MLLM usage experience. The description of each psychotherapist and the corresponding profile can be found in Table \ref{t1}. Particularly, as previous studies pointed out that professional experience might contribute to varied attitudes to computerized psychotherapy \citep{mcdonnell2013attitudes}, we expect to further compare and analyze the similarities and differences of views from groups of psychotherapists with different levels of working experience. Thus, we divided them into three groups according to their years of experience: 3-5 years (Junior therapist group, J), 5-10 years (Mid-career therapist group, M), and more than 10 years (Senior therapist group, S) \citep{sun2024understanding}. Each group consisted of 5 participants. Moreover, referring to their self-reported therapeutic orientation, to eliminate the difference between therapy styles \citep{bagaric2020measuring} and extract consensual within the group, we tried to balance the psychotherapists' distribution of therapeutic orientation within each group.

\begin{table}[h]\centering
\caption{Demographic Information of the Interviewed Students.}
\label{t2}
\small
\begin{tabular}{c|c|c|c}
\toprule
ID & Age& Gender & Major \\
\hline
I1 & 20  & Female & Histrory                        \\
I2 & 22  & Female & Popular Music                   \\
I3 & 23  & Female & Applied Mathematics             \\
I4 & 28  & Female & Applied Psychology              \\
I5 & 18  & Male   & Geosciences                     \\
\hline
D1 & 21  & Male   & Computer Science                \\
D2 & 23  & Female & Mechanics                       \\
D3 & 23  & Female & Mental Health Education         \\
D4 & 20  & Female & Journalism                      \\
D5 & 21  & Male   & Computer Science                \\
\hline
O1 & 27  & Male   & Political Science               \\
O2 & 19  & Male   & Electronic Sciences             \\
O3 & 21  & Female & Applied Psychology              \\
O4 & 20  & Female & Chinese Literature              \\
O5 & 22  & Female & Radio and Television Studies    \\
\hline
N1 & 21  & Male   & Industrial Engineering          \\
N2 & 21  & Male   & Electronic Information Science  \\
N3 & 23  & Female & Transport Engineering           \\
N4 & 20  & Female & Industrial Engineering          \\
N5 & 21  & Male   & Statistic       \\
\bottomrule

\end{tabular}
\textnormal{\\Notes: 'I' means the mental illness group; 'D' stands for mental distress group; 'O' means mental disorder group; and 'N' is normal group.}
\end{table}

\subsubsection{Study II: Semi-structured Interview with College Students}
In total, 20 college students (8 males and 12 females) in mainland China were involved in our semi-structured interviews. All participants were currently enrolled college students (from bachelor students to Ph.D. students) and had MLLM usage experience. Their demographic information is presented in Table \ref{t2}. Similarly, as preliminary research found that people's mental health level is one of the dominant factors in their attitudes toward AI tools \citep{he2024influencetaskgroupdisparities}, we divided them into four groups based on the severity of their psychological issues: the normal group (N), and three groups with different levels of mental health issues (i.e., mental distress group (D), mental disorder group (O), and mental illness group (I)). Each group consisted of 5 participants, allowing us to explore how the severity of mental issues influences users' views on the application of MLLMs in psychotherapy. 

Participants in the normal group were recruited through on-campus posters, while others were invited by their psychotherapists and chose to participate voluntarily in our study. For group classification, we selected individuals in the normal group who had not participated in any form of psychotherapy. For the other groups, we consulted four senior psychotherapists to discuss the categorization of participants before starting the interviews. Each participant was classified based on the severity of their mental health issues, as detailed in their past counseling reports. These reports were provided by the participants’ own therapists from the group of four. Importantly, to prevent any potential stigma and the challenges that students might face, the grouping process was anonymized in the counseling reports.

\subsubsection{Study III: Survey with College Students}
An online survey study was conducted, and participants were recruited from online forums. Based on the in-depth interview with 20 college students, this study aims to quantitatively validate findings from Study II, and further investigate influential socio-psychology factors toward users' usage intention of MLLM for psychotherapy in different scenarios. Thus, the target population is current college students who have attended campus psychotherapy. We asked all participants to submit relevant materials to support their psychotherapy experience (e.g., Screenshots of service appointments, diagnostic reports.). After our manual screening, all materials submitted that would reveal personal information were removed. A total of 422 participants completed the questionnaire. We excluded responses from those who had never heard LLMs or taken campus psychotherapy before, and screened the remaining answers based on three quality-checking questions (i.e., "If you are answering the question carefully, please select the second/third option"). Ultimately, 366 participants (198 female and 168 male) with an average age of 21.0 years old (minimum: 18, maximum: 29, standard deviation: 2.2) were included for analysis. They were each compensated 10 RMB for their time spent completing the 10-minute questionnaire.

\subsection{Procedures of Human Therapeutic Alliance Interview}
\subsubsection{Study I: Focus Group Interview with Psychotherapists.} Each focus group session lasted approximately 90 minutes and was moderated by two experienced researchers: one with a background in computer science and the other in psychology. The focus group study was held from April to May of 2024 via the online video meeting tool, the Tecent Meeting\footnote{https://voovmeeting.com/}. Upon entering the virtual meeting room, we notified them about the details of the interview, and obtained their consent and demographic information through an online questionnaire. Then, moderators introduced the purpose of the study, explained the discussion process, and emphasized confidentiality and their right to withdraw at any time. Next, a short video introducing MLLM and its functions that might be related to psychotherapy (e.g., multi-modal emotion perception, emotional soothing) was played before the interview. 

After the interview began, each participant was asked to introduce themselves to others briefly. Then, participants were first asked about their attitudes toward using MLLMs in campus psychotherapy. This was followed by discussions about how to divide campus psychotherapy into different phases, and which specific phases the LLMs could be applied, particularly in addressing common mental health issues in universities. As there is no therapist in daily counseling and autonomous therapy theoretically, we did not design specific questions for these two types of MLLM. We then queried their opinions on essential features and functions that can make MLLMs effective in campus psychotherapy. Finally, participants gave their own opinions about potential challenges and concerns related to the use of MLLMs. The whole interview was audio-recorded, transcribed, and supplemented with field notes to capture non-verbal cues and group dynamics. The outline of the focus group interview is in \ref{app:A}.

\subsubsection{Study II: Semi-structured Interview with College Students.} In this study, a 30-minute semi-structured interview was conducted with each participant. This study was finished in June 2024. Similar to Study I, all participants were informed of the content of our study and gave their consent. All interviews were held and recorded online. In particular, the same video about MLLM was provided to all participants. While the general structure mirrored Study I, the specific questions discussed were tailored to the patients. Specifically, after we obtained the information regarding the consensual phases of psychotherapy from therapists, we investigated student users' attitudes, expected functions, and concerns about using MLLM in different phases, for two types of psychotherapy without therapists (i.e., daily counseling and autonomous therapy). The questions used in the semi-structured interview are shown in the \ref{app:B}.

\begin{sidewaystable}[thp]
    \centering
\caption{Descriptive Statistics of the Factors of Interest.}
\label{t3}
\scriptsize
    \begin{tabular}{l|l|l}
\toprule
        \textbf{Variable} & \textbf{Description} & \textbf{Distribution of extracted variables} \\ \hline

        Age & Age of participants. & Mean: 21.0 (SD: 2.2, min: 18, max: 29) \\ \hline
        
        \multirow{2}{*}{Gender} & \multirow{2}{*}{Gender of participants.} & Male (n=168, 45.9\%) \\ 
        ~ & ~ & Female (n=198, 54.1\%) \\ \hline
        
        \multirow{2}{*}{Education} &  \multirow{2}{*}{The education level of participants.} & Below bachelor’s degree (n=243, 66.4\%) \\ 
        ~ & ~ & Bachelor’s degree and above (n=123, 33.6\%) \\ \hline
        
         \multirow{3}{*}{Usage Frequency} &  \multirow{3}{*}{How frequency participants used MLLMs.} & Rarely (n=133, 36.3\%) \\ 
        ~ & ~ & Sometimes (n=123, 33.6\%) \\ 
        ~ & ~ & Always (n=110, 30.1\%) \\ \hline

        \multirow{3}{*}{Anxiety} & \multirow{6}{*}{\shortstack{Seven questions from Hospital Anxiety and Depression Scale (HADS)\\\citep{zigmond1983hospital} for anxiety and depression each. Each\\question has four choices as coded from 0 to 3. \\- 0-7: Normal; - 8-10: Borderline abnormal \\(borderline case);- 11-21: Abnormal (case)}} &  Normal (n=28, 6.6\%)  \\
~ & ~ & Borderline abnormal (n=109, 29.8\%) \\ 
        ~ & ~ & Abnormal (n=229, 62.6\%) \\
        \cline{1-1} \cline{3-3}
        \multirow{3}{*}{Depression} & ~ &  \multirow{3}{*}{\shortstack[l]{Normal (n=55, 14.5\%)\\Borderline abnormal (n=194, 53.0\%)\\Abnormal (n=117, 32.5\%)}}\\
        ~ & ~ &  \\ 
        ~ & ~ & \\ 
        \hline

        \multirow{4}{*}{Social Identity} &  \multirow{4}{*}{\shortstack{We selected three positive items and one negative item from\\\citep{feitosa2012social} to measure participants'  social identity.\\ As scores rose, participants identified themselves more as\\ members of the human race.}} & \multirow{4}{*}{Mean: 14.5 (SD: 3.9, min: 4, max: 20)} \\  & & \\  & &  \\& &   \\ \hline

        \multirow{2}{*}{Privacy Awareness} &  \multirow{2}{*}{\shortstack{The sum of four measurements in \citep{stewart2002empirical},\\regarding users' privacy concerns to MLLMs in psychotherapy. }} & \multirow{2}{*}{Mean: 14.1 (SD: 3.0, min: 5, max: 20)} \\  & &   \\ \hline

        \multirow{2}{*}{Trust in MLLM} &  \multirow{2}{*}{\shortstack{The Five-item Trustworthiness scale (FIFT) \citep{franke2015advancing}\\ regarding users’ trust in MLLMs.}} & \multirow{2}{*}{Mean: 19.8 (SD: 2.8, min: 13, max: 25)} \\  & &   \\ \hline

        \multirow{3}{*}{Relative Status} &  \multirow{3}{*}{\shortstack{To measure the manipulation of two social roles of MLLM users \\ perceived, we used the three items from \citep{zhang2023tools}. \\As scores rise, MLLM is seen as a mentor rather than a servant.  }} & \multirow{3}{*}{Mean: 7.7 (SD: 1.6, min: 2, max: 10)} \\  & &   \\& &   \\ \hline

        \multirow{3}{*}{Anthropomorphism} &  \multirow{3}{*}{\shortstack{The user perceived anthropomorphism level of MLLM in \\ psychotherapy was assessed by three-items in \citep{zhang2023tools}.\\ As scores rise, the more anthropomorphic MLLM is. }} & \multirow{3}{*}{Mean: 10.8 (SD: 2.4, min: 3, max: 15)} \\  & &   \\& &   \\ \hline

        \multirow{7}{*}{\shortstack{Acceptance to \\use MLLM}} & \multirow{7}{*}{\shortstack{Participants' acceptance of using MLLM in five phases assessed by the \\mean scores of five items \citep{adell2010acceptance}:\\ Acceptance1: the information gathering phase \\ Acceptance2: the working phase\\ Acceptance3: the feedback and evaluation phase\\ Acceptance4: daily counseling\\ Acceptance5: autonomous therapy}} & \multirow{7}{*}{\shortstack{Acceptance1: Mean: 4.1 (SD: 0.5, min: 1, max: 5)\\Acceptance2: Mean: 4.0 (SD: 0.6, min: 1, max: 5)\\Acceptance3: Mean: 4.2 (SD: 0.6, min: 1, max: 5)\\Acceptance4: Mean: 3.9 (SD: 0.6, min: 1, max: 5)\\Acceptance5: Mean: 3.8 (SD: 0.6, min: 1, max: 5)}} \\  & &   \\
 & &   \\  & &   \\  & &   \\ & &   \\  & &   \\ 
 \bottomrule     
    \end{tabular}

\end{sidewaystable}

\subsubsection{Design of Survey for College Students}
To supplement our findings from student interviews, we conducted a survey study on user attitudes towards MLLMs in psychotherapy. This involved an online questionnaire, outlined in Table \ref{t3}, which captured both demographic information and user perspectives. Specifically, Q1 to Q3 gathered demographic information (i.e., age, gender, education) and assessed respondents' prior experience with MLLMs and psychotherapy, which may influence user attitudes \citep{wang2024evaluating, wang2024young}. Q4 measured participants' familiarity with popular LLMs (e.g., ChatGPT), as there were established relationships among user knowledge, trust, and acceptance of automation \citep{wang2024trust}. To categorize participants into mental health groups corresponding to Study II, Q5 utilized the Hospital Anxiety and Depression Scale (HADS) \citep{zigmond1983hospital}, identifying three levels of mental issues. It is worth noting that, as those who never experienced psychotherapy can hardly imagine the procedure of psychotherapy with the descriptions provided in survey only, we abandoned the samples from participants who were classified into the normal level in both anxiety and depression. Drawing on insights from Study II, Q6 to 10 explored additional factors that may potentially impact user attitudes toward MLLM tools (i.e., their own social identity \citep{feitosa2012social} and privacy protection awareness \citep{yoo2023discussing}, their trust \citep{wang2024young}, relative status, and anthropomorphism \citep{zhang2023tools} to MLLM). Finally, Q11 (Acceptance1 to 5), measured user willingness to use MLLMs in five phases of psychotherapy. 

\subsection{Analysis and Coding Methods}

\subsubsection{Focus Group and Semi-Interview Analysis}
This work employed a mixed-methods approach, combining qualitative and quantitative analysis, to examine user perspectives obtained in focus groups (Study I) and semi-structured interviews (Study II). The data was analyzed using a rigorous thematic analysis procedure. The process began with transcribing audio recordings from Tencent Meeting into texts, ensuring accuracy through careful calibration with the original audio. Two researchers independently reviewed the transcripts and employed a combination of deductive and inductive coding techniques. Firstly, drawing on the research questions and interview guide, four overarching themes were established: attitude, scenario, function, and challenges. The researchers then independently coded segments of the interview data related to these themes. Simultaneously, we engaged in open coding, identifying key concepts and assigning descriptive labels to relevant text portions. Through iterative discussions, we reconciled the codings, merged frequently occurring labels, and developed a comprehensive codebook. This codebook was further refined through discussions with the broader research team. To further guarantee the quality of the analysis, a third researcher independently reviewed the coding framework. Quantitative analysis was also conducted to examine the frequency, distribution, and relationships between codes. Finally, the research team synthesized findings from both the qualitative and quantitative analyses, drawing connections between the themes and exploring key patterns. It is important to note that the semi-structured nature of the interviews allowed for flexibility in questioning. Consequently, not all participants responded to the same set of questions. Therefore, the analysis focused on the specific subset of participants who provided relevant data for each code and theme, rather than the entire pool of 30 interviewees.

\subsubsection{Survey Analysis}
For the quantitative analysis, we used SAS OnDemand for Academics to quantify the effects of participants' backgrounds and their socio-psychological factors on users' attitudes towards using MLLM in psychotherapy. Mixed linear regression models (using Proc MIXED) were built for five continuous dependent variables (Acceptance1 to 5). All demographic factors, five socio-psychological factors, and their two-way interactions were used as the independent variables in the initial full models. We adopted backward stepwise selection procedures based on model fitting criteria, and used Variance Inflation Factor (VIF) to mitigate the issue of multicollinearity. The Tukey-Kramer post-hoc tests were conducted for significant independent variables (p$<$.05).

\section{Results}

\subsection{Results from Study I}

\begin{figure*}
\begin{center}
\includegraphics[scale=0.33]{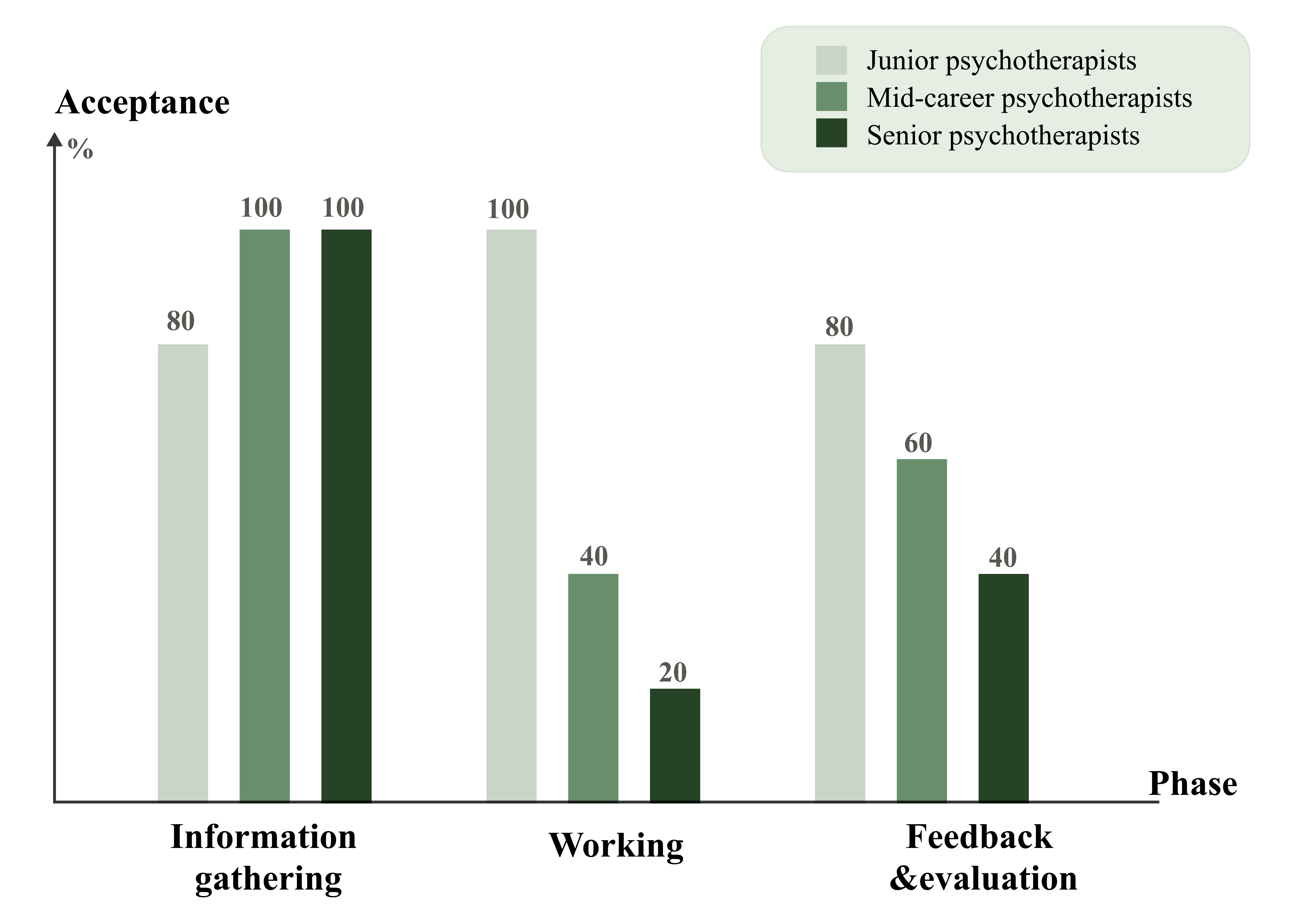}\\
\end{center}
\vspace{-4mm}
\caption{Statistics about psychotherapists' views of MLLM application scenarios in campus psychotherapy.}\label{f1}
\vspace{-4mm}
\end{figure*}

\subsubsection{Application phases} As shown in Figure \ref{f1}, in the information gathering phase of psychotherapy, most junior psychotherapists recognized the potential of MLLMs. Junior therapist J4 remarked, \textit{"The first step in information gathering is to understand the client’s family background and developmental history... I personally think this step can be supported by MLLMs."} Mid-career and senior psychotherapists also agreed on the suitability of MLLMs for this phase. Mid-career therapist M5 noted, \textit{"After collecting the basic questions and information, MLLMs can match the client with a suitable psychotherapist... I think this approach can significantly reduce the initial investment of manpower and resources."} Senior therapist S1 suggested, \textit{"When students are waiting for an appointment, they might fill out a form... they could be guided to download and complete an assessment themselves while in the waiting room."}

During the working phase, junior psychotherapists agreed on the applicability of MLLMs. J4 suggested, \textit{"During the working phase, I think MLLMs can help with the unidirectional collection of repeated information… The second aspect is providing prompts... MLLMs might help enhance the completeness or fluency of my output."} However, fewer psychotherapists in mid-career and senior groups recognized the applicability of MLLMs in this phase. For example, senior therapist S3 highlighted, \textit{"If MLLMs could be employed in crisis assessment, it would significantly reduce the pressure on psychotherapists."}

In the feedback and evaluation phase, most junior therapists indicated they would use MLLMs for support. Junior therapist L3 remarked, \textit{"During the termination phase, clients might also provide MLLMs with feedback on their progress... I think it’s important to observe from multiple perspectives to understand how well their action plan has been executed."} Among mid-career therapists, more than half believed MLLMs are helpful, with mid-career therapist M2 stating, \textit{"In the final termination phase, dealing with common termination procedures and evaluating and providing feedback (with MLLMs) ... is, in my opinion, appropriate."} However, only a few senior therapists discussed the applicability of MLLMs in this phase.

\subsubsection{Expected Function} In general, psychotherapists emphasized the importance of robust triage functionalities in MLLMs, which involves initial assessment and categorization of a client’s symptoms. Junior therapist J2 noted, \textit{"I believe that for MLLMs to effectively identify such situations, they must classify the clients accordingly... If the client suffers from conditions that might blur their subjective awareness and objective reality, an additional follow-up questioning phase is required."} Additionally, the triage function can assist psychotherapists by providing initial recommendations for further inquiry or intervention, helping to ensure that clients receive the most appropriate care based on their specific needs. S4 further elaborated, \textit{"When a client is unable to distinguish between subjective and objective aspects of a particular issue... it would be beneficial to ask additional questions regarding general knowledge."}

Besides, therapists with different experience levels all expressed a strong desire for MLLMs to incorporate emotional support functionalities. Junior therapist J3 suggested, \textit{"I believe that more precise and varied forms of emotional support could better assist the counseling process, making it more akin to emotional exchanges between individuals."} Similarly, in the Mid-Career group, M4 mentioned, \textit{"For example, turning on a camera could provide a sense of interaction."} Additionally, there were two therapists expressed a preference for music options. For instance, S2 remarked, \textit{" When a client is excessively anxious and tense... MLLMs could employ relaxation training techniques to substitute the psychotherapist... These techniques could include small methods for emotional stabilization."}

Psychotherapists highlighted the need for MLLMs to include monitoring and suggestion functions to assist in the therapeutic process. J1 stated, \textit{"MLLMs can assist psychotherapists by monitoring the counseling process, recording the content of sessions, and capturing various non-verbal cues... such as gestures, tone, and facial expressions."} M3 also emphasized the importance of enabling MLLMs to assess whether the current situation is suitable for further intervention and determine the severity of the issue. She noted, \textit{"As a psychotherapist if I wish to delegate some tasks... it would be very helpful if MLLMs could accurately record the key points of each session and even provide suggestions for the next stage."} They also mentioned that theoretical guidance could be provided in these suggestions. For example, M2 mentioned, \textit{"I hope MLLMs can offer robust models to guide the counseling process."} This perspective was also supported by more than half of senior psychotherapists, with S2 stating, \textit{"The ideal tool would offer a wealth of theoretical frameworks... This would provide valuable reference material for me. I hope it has a retrieval function."} The other two therapists also expressed their support for this functionality.

\subsubsection{Concerns} Psychotherapists widely expressed concerns about potential breaches of client data confidentiality when using MLLMs, especially in online therapy environments where the risk of data leakage is higher. Across all three groups, half of psychotherapists highlighted this issue. For example, in the senior group, S3 stated, \textit{"This violates the ethical principles of psychotherapy... Our confidentiality protocols, including recording sessions, require the client’s explicit consent."} They were also concerned about whether MLLMs could effectively manage transference (i.e., a psychological phenomenon where an individual unconsciously redirects feelings and expectations from past relationships onto a current relationship, often seen in therapy). In the junior group, some therapists questioned this ability, and in the mid-career group, a few therapists expressed similar concerns. As junior therapist J2 questioned, \textit{"If an MLLM is used to provide psychotherapy through an app, and the client develops transference towards the MLLM, how should this be addressed? ... MLLMs may not necessarily recognize when transference occurs."}

Furthermore, therapists believe that MLLMs may struggle to establish deep emotional connections with clients. Junior therapist J1 stated, \textit{"My concern is that humans will never be able to establish a true emotional connection with machines."} Most of the interviewees in the mid and senior-career groups mentioned this concern. S3 mentioned, \textit{"In psychotherapy, much of what we do involves the warm emotional flow between humans. An MLLM may not be able to provide such an emotional flow..."} Concerns were also raised about the lack of flexibility in MLLMs, as junior therapist J3 expressed, \textit{"I am concerned that MLLMs may offer similar advice to many clients, lacking the ability to provide more individualized and targeted guidance."} S1 in the senior group also noted, \textit{"The advice provided by MLLMs may not always be as precisely targeted as needed."}

\subsection{Results from Study II}

\begin{figure*}
\begin{center}
\includegraphics[scale=0.22]{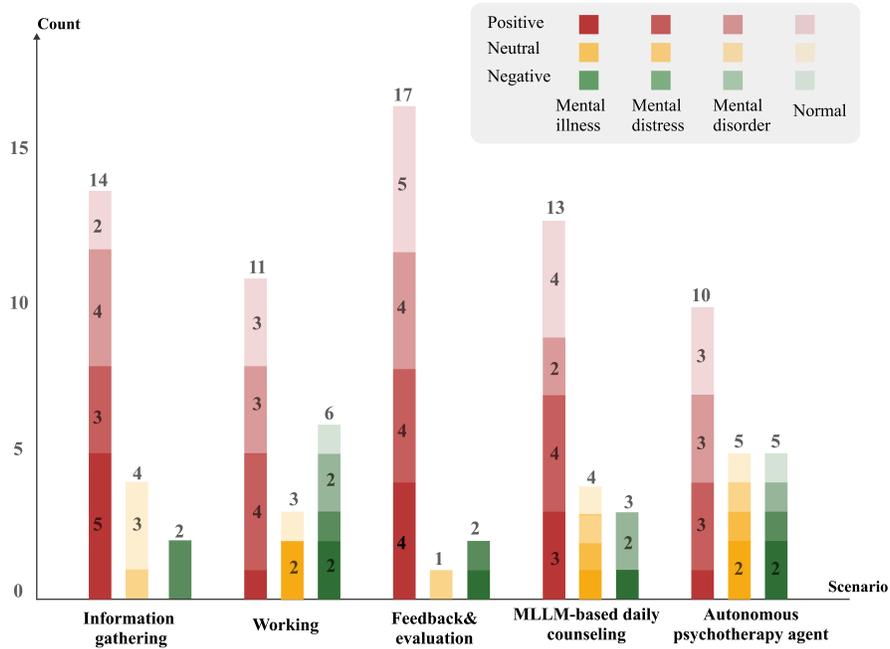}\\
\end{center}
\vspace{-4mm}
\caption{Quantitative data about student clients' attitudes from the survey study.}\label{f2}
\vspace{-4mm}
\end{figure*}

\subsubsection{Attitudes}
In the context of MLLM-assisted face-to-face psychotherapy, referring to the consensual of interviewed therapists, the counseling process can be generally divided into three phases: the information-gathering phase, the working phase, and the feedback and evaluation phase. We investigated students’ attitudes toward the integration of MLLMs in each phase, and the results are in Figure \ref{f2}. In the information-gathering phase, almost all of the students expressed a willingness (including positive and neutral) to accept the use of MLLMs by therapists for collecting and analyzing personal information, although a few had concerns. In the working phase, acceptance slightly decreased, including 3 still holding conditional acceptance (neutral). In the feedback and evaluation phase, still almost all of the students supported using MLLMs to summarize and analyze the consultation process, with only one student expressing neutral attitudes. Overall, students demonstrated a positive attitude toward MLLM assistance, with high acceptance levels.

Regarding MLLM-based daily counseling, more than half of the students held a positive attitude, while a few indicated conditional acceptance, and three students maintained a negative stance. When considering autonomous psychotherapy agents, only half students showed a positive attitude, while a few indicated conditional acceptance, and less than a half explicitly opposed the replacement of human therapists.

When comparing different groups, the mental illness group exhibited the highest acceptance in the information-gathering phase, followed by the mental disorder group. In the mental distress group, some students were unable to accept MLLM-assisted therapy, and only a few in the normal group expressed full acceptance. In the working phase, the mental illness group had the lowest acceptance, whereas most students in the normal group maintained a positive attitude toward AI-assisted therapy. In the feedback and evaluation phase, all students in the normal group showed a support perspective, with only one student from the mental disorder group expressing concerns. Particularly, in daily counseling, the normal and mental distress groups had the highest acceptance, with no students opposing it, while a few opposing opinions were noted in the mental disorder and mental illness groups. In the case of autonomous psychotherapy agents, the mental illness group showed higher opposition, with only a few expressing a full support, while the other three groups generally maintained a supportive stance.

\begin{figure*}
\begin{center}
\includegraphics[scale=0.08]{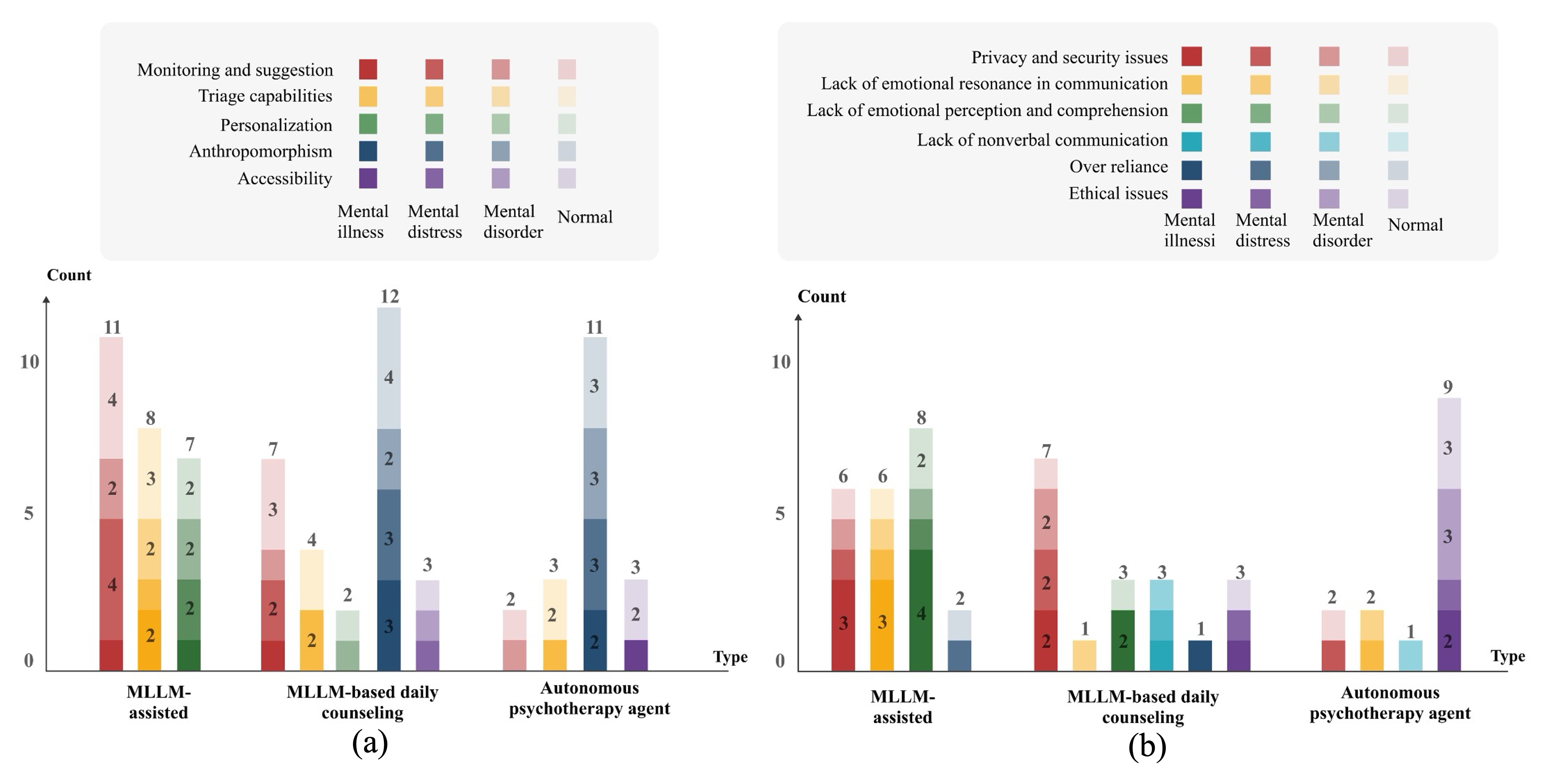}\\
\end{center}
\vspace{-4mm}
\caption{Quantitative data about student clients' expected functions and concerns in the interview.}\label{f3}
\vspace{-4mm}
\end{figure*}

\subsubsection{Expectation of MLMM Functions}
The results are visualized in Figure \ref{f3}(a). The essential functions of MLLM-assisted face-to-face psychotherapy were categorized into three themes: 1. Monitoring and suggestion; 2. Triage capabilities; 3. Personalization. Specifically, monitoring and suggestions were highlighted by half of the students. For example, N2 commented, \textit{"I would build my trust only when it can provide real-time and accurate feedback on the emotions conveyed through speech and expressions."} Similarly, in the mental distress group, D3 stated, \textit{"From a technical perspective, as long as it can accurately capture my state, possess relevant knowledge, and provide correct feedback to the therapist, I think that would suffice."} O4 from the mental disorder group also said, \textit{"The facial expression recognition feature seems quite useful to me."} Notably, one student also mentioned the importance of crisis detection. N3 remarked, \textit{"What if someone is suffering from a disorder, and they suddenly attempt to attack or something? Could the MLLM foresee and alert beforehand? For those requiring security in emergencies, this would be vital."}

Moreover, less than half of the students anticipated that such MLLM would possess triage capabilities focusing on information gathering, analysis, and referral. The first point is information collection, a point raised by students from both the mental distress and mental disorder groups. D3 said: \textit{"For instance, MLLM could assist the therapist in gathering information during or before the counseling."} O3 added: \textit{"I feel that in human conversations, some details might be overlooked, but MLLM could be more comprehensive."} The next point is analysis and assessment, as mentioned by N4 from the normal group, \textit{"In psychotherapy, it would involve an assessment of the person’s overall condition."} Then, classification and referral, as noted by I4: \textit{"When you first arrive, the MLLM could help with triage and assign you to the most suitable therapist, or one specialized in your concerns."} Some other students highlighted the significance of MLLM-assisted personalization, where the system would adjust its services based on the patient’s needs. In the mental illness group, one student suggested the idea of training AI to enhance personalization. I4 stated, \textit{"I wonder if the MLLM needs to be trained to better suit the user."} Similarly, others mentioned tailoring services according to the patient’s specific needs. O2 noted, \textit{"It should be able to provide services based on the patient’s requirements."} Additionally, archiving and memory functions were also considered important. D2 mentioned, \textit{"The patient could communicate with the MLLM, and it would store the conversation. Then, for the next session, you could pick up where you left off."}

In terms of MLLM-based daily counseling, the expected features include monitoring and suggestion, and the memory function under personalization. Seven students expressed interest in the monitoring and suggestion feature. Additionally, students highlighted two new features for daily counseling MLLM: anthropomorphism and accessibility. Over half of the students felt that anthropomorphism was crucial for an MLLM in this context. D3 added, \textit{"If I could receive the emotional responses I want from the MLLM, that would greatly improve my experience."} Furthermore, in terms of the MLLM’s persona, three students expressed a desire to customize it. O5 mentioned: \textit{"It would be beneficial if we can choose the MLLM’s persona."} Interestingly, most students in the mental illness group prioritized whether the MLLM had human-like characteristics. As for the accessibility feature, students from the normal, mental distress, and disorder groups expressed an expectation for the MLLM to be easily accessible and user-friendly. N3 said: \textit{"If it’s too complicated, I won’t bother setting it up."} Students also wanted an easily usable and portable interface, as noted by D5 and O4. Finally, regarding an autonomous psychotherapy agent, over half of the students reiterated the importance of anthropomorphism, desiring human-like appearances and emotions. Two students mentioned the need for learning and summarizing capabilities, and physical touch functions were also suggested. D4 added: \textit{"If I’m in an emotional breakdown, I might want a hug... some kind of physical contact for comfort."}

\subsubsection{Concerns}
The concerns expressed by interviewed students about three types of psychotherapy (MLLM-assisted, daily counseling, and autonomous therapy) can be categorized into six themes: 1. Insufficient emotional perception and comprehension; 2. Insufficient emotional support in communication; 3. Privacy security issues; 4. Ethical issues; 5. Lack of nonverbal communication; 6. Over-reliance. Descriptive statistics data regarding these concerns is shown in Figure \ref{f3}(b).

First, insufficient emotional perception and comprehension were the most commonly expressed concerns across all three types of psychotherapy. In both MLLM-assisted face-to-face psychotherapy and MLLM-based daily counseling, students were particularly worried about the MLLM’s inability to accurately interpret emotions. I1 highlighted: \textit{"MLLM cannot perceive human facial expressions or emotional fluctuations."} In autonomous therapy, students felt MLLMs lacked a deep understanding, with O5 stating, \textit{"There are many abstract, profound emotions that machines simply cannot replicate."} These concerns were more prominent in the mental illness group. Second, privacy security issues were another prevalent concern among students, spanning across all three types of MLLM. O1 expressed worry about potential leaks of personal information. A notable trend emerged in the mental illness group, where concerns about privacy declined from MLLM-assisted face-to-face psychotherapy to autonomous psychotherapy agents.
In addition, students across all three types of MLM expressed concerns about insufficient emotional support in communication, especially in the context of MLLM-assisted face-to-face psychotherapy. Specifically, students questioned MLLM’s ability to provide rich, emotionally nuanced language. For instance, O4 asked: \textit{"Can MLLM deliver richness and variety of language styles, such as humor, I enjoy?"} and O2 was skeptical about MLLM simulating emotionally guided counseling effects. Concerns about real-life experience were also noted in autonomous therapy, with D1 remarking, \textit{"It cannot offer advice based on real experiences."} Ethical issues were of particular concern in the context of daily counseling and autonomous psychotherapy agents, with a clear upward trend compared to MLLM-assisted therapy. Students worried that highly advanced MLLM could manipulate them verbally or even mentally during psychotherapy. O5 highlighted the risks of MLLM miscommunication affecting vulnerable individuals. Similarly, N3 raised the issue: \textit{"If MLLM becomes uncontrollable, it might exploit human weaknesses. If I had psychological issues, I could be easily influenced."}
Two other concerns were raised by a minority of students: Over-reliance and lack of nonverbal communication, which may reduce the effectiveness of therapy. O2 commented: \textit{"This over-reliance is mutual. First, if therapists overly rely on MLLM, psychotherapy may lose communication between humans. At the same time, patients might over-rely on MLLM, leading to self-isolation."} Lack of nonverbal communication was mostly raised in the latter two types (daily counseling and autonomous agent). Three students were concerned that MLLM, lacking a physical presence, would be unable to provide important non-verbal forms of support in therapy, such as physical touch. D2 emphasized: \textit{"No matter how advanced it becomes, the lack of physical interaction is a limitation."}

\begin{table}[H]\centering
\caption{Summary of Statistical Results.}
\label{t4}
\scriptsize
\begin{tabularx}{\textwidth}{c|X|c|X|c}
\toprule
\textbf{DV} & \textbf{IV}    & \textbf{F-value }          & \textbf{Estimation [95\% CI]} & \textbf{\textit{p}} \\
\hline
\multirow{4}{*}{Acceptance1}        
& Relative Status             & F(1, 359) = 63.72 & 0.11 [0.08, 0.14]   & $<$.0001$^{*}$              \\
& Trust in MLLM                     & F(1, 359) = 17.97   & 0.01 [0.00, 0.02] &$<$.0001$^{*}$             \\
& Usage Frequency                     & F(2, 359) = 10.20   & - &$<$.0001$^{*}$            \\
&  Trust in MLLM * Usage Frequency                    & F(2, 359) = 9.67 & -   &$<$.0001$^{*}$               \\
                        
\hline
\multirow{2}{*}{Acceptance2}           
&       Relative Status             & F(1, 361) = 43.32 & 0.11 [0.07, 0.14]   & $<$.0001$^{*}$              \\
& Trust in MLLM                     & F(1, 361) = 40.94   & 0.07 [0.05, 0.10] &$<$.0001$^{*}$             \\
& Anxiety                     & F(2, 361) = 8.85   & - &.0002$^{*}$            \\

\hline
\multirow{4}{*}{Acceptance3} 
& Relative Status             & F(1, 354) = 23.70 & 0.08 [0.05, 0.11]   & $<$.0001$^{*}$              \\
& Trust in MLLM                     & F(1, 354) = 24.01   & 0.004 [-0.04, 0.05] &$<$.0001$^{*}$             \\      
& Usage Frequency                     & F(2, 354) = 7.12   & - &.001$^{*}$             \\
& Trust in MLLM * Usage Frequency                    & F(2, 354) = 7.49   & - &.0007$^{*}$             \\
& Anxiety                     & F(2, 354) = 2.42   & - &.002$^{*}$             \\
& Social Identity                    & F(1, 354) = 1.04   & -0.01 [-0.03, 0.01] &.3             \\ 
& Social Identity * Anxiety                    & F(2, 354) = 3.71   & - &.03$^{*}$             \\
\hline

\multirow{2}{*}{Acceptance4}           
&  Trust in MLLM                     & F(1, 363) = 84.50   & 0.09 [0.07, 0.11] &$<$.0001$^{*}$             \\
& Anthropomorphism                    & F(1, 363) = 13.21   & 0.05 [0.02, 0.07] &.0005$^{*}$              \\
\hline
\multirow{2}{*}{Acceptance5}           
&  Trust in MLLM                     & F(1, 362) = 70.92   & 0.09 [0.07, 0.11] &$<$.0001$^{*}$             \\
& Anthropomorphism                    & F(1, 362) = 17.90   & 0.06 [0.03, 0.09] &$<$.0001$^{*}$              \\
& Privacy Awareness                    & F(1, 362) = 12.60   & 0.04 [0.02, 0.06] &.0005$^{*}$              \\
\bottomrule

\end{tabularx}
\textnormal{\\Notes: In this table and the following tables, $^{*}$ marks significant results (\textit{p}$<$.05), DV indicates the dependent variable and IV is the independent variable.}
\end{table}

\begin{table}[H]\centering
\caption{Significant Post-hoc Results for Discrete Independent Variables.}
\label{t5}
\scriptsize
\begin{tabular}{c|c|c|c|c|c|c}
\toprule
\textbf{DV}    & \textbf{IV}                         & \textbf{Level}                    & \textbf{Compared level}             & \textbf{Estimation[95\% CI]}   & \textbf{t value}     & \textbf{\textit{p}} \\
\hline
\multirow{2}{*}{Acceptance1}  & \multirow{2}{*}{Usage Frequency}                 & Rarely                          & Always                                & -0.19 [-0.34, -0.03]      & t(359)=-2.82  & .01$^{*}$              \\

     &                 & Sometimes                          & Always                                & -0.17 [-0.32, -0.01]      & t(359)=-2.47  & .04$^{*}$              \\
\hline
Acceptance2  & Anxiety                & Borderline abnormal                           & Abnormal                                 & 0.27 [0.12, 0.42]      & t(361)=4.19  & .0001$^{*}$              \\
\hline
Acceptance3  & Usage Frequency                & Rarely                          & Always                                & -0.09 [-0.17, -0.01]      & t(354)=-1.20  & .047$^{*}$              \\
\bottomrule

\end{tabular}
\textnormal{\\Notes: Estimate is the difference between IV level and IV level compared to.}
\end{table}

\subsection{Results from Study III}

\begin{figure*}
\begin{center}
\includegraphics[scale=0.26]{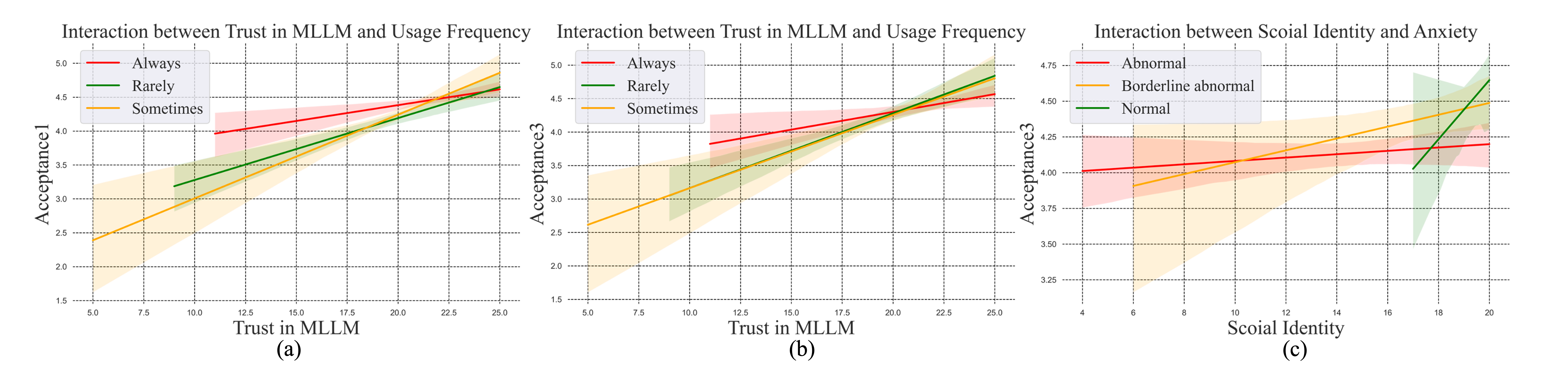}\\
\end{center}
\vspace{-4mm}
\caption{Visualization of significant interaction effects. Each figure illustrates the linear relationship after fitting the sample distribution with the best-fitted models in Table \ref{t2}, as well as the 95\% confidence interval (CI). Subfigure (a) corresponds to the model of Acceptance1, (b) and (c) is of the model of Acceptance3.}\label{f5}
\vspace{-4mm}
\end{figure*}

The results of the survey study are shown in Table \ref{t4}, \ref{t5}, and significant two-way interactions are visualized in Figure \ref{f5}. These results reveal the relationships between various independent variables and participants' acceptance of MLLMs in different psychotherapy scenarios.

Firstly, the Acceptance1, which involves using MLLMs to assist human therapists before face-to-face counseling, was associated with several variables. Specifically, the Relative Status and Trust in MLLM had a significant positive effect on Acceptance1. Additionally, Usage Frequency was also a significant predictor of Acceptance1 and had a significant interaction effect with Trust in MLLM and Usage Frequency. Referring to Figure \ref{f5}(a), we found that the effect of Trust in MLLM on Acceptance1 appears stronger for those who used MLLMs more frequently. At the same time, Relative Status and Trust in MLLM were positively associated with Acceptance2, which involves using MLLMs during face-to-face counseling. Further, those with borderline abnormal anxiety presented significantly higher acceptance than those with abnormal anxiety levels. In the case of Acceptance3, involving MLLMs after face-to-face counseling, similar to the Acceptance1 model, the significant effects of Relative Status, Trust in MLLM and its interaction with Usage Frequency were observed again. Moreover, a significant interaction between Social Identity and Anxiety was found. As shown in Figure \ref{f5}(c), we first found that the Social Identity score of those in the normal group was centralized in the higher interval than those with anxiety issues. Besides, apart from the positive association between Social Identity and Acceptance3, we found the association was weaker when the Anxiety level got severe.

In addition to the first three models corresponding to MLLM-assisted face-to-face counseling, for students' acceptance towards using MLLMs in daily therapy (Acceptance4), we identified that Trust in MLLM and Anthropomorphism were positively associated with Acceptance4. Finally, Acceptance5, which indicates the acceptance of independent psychotherapy by MLLMs, was associated with Trust in MLLM, Anthropomorphism, and privacy awareness.

\section{Discussion}
\subsection{Attitudes from human therapy alliance}
\begin{figure*}
\begin{center}
\includegraphics[scale=0.5]{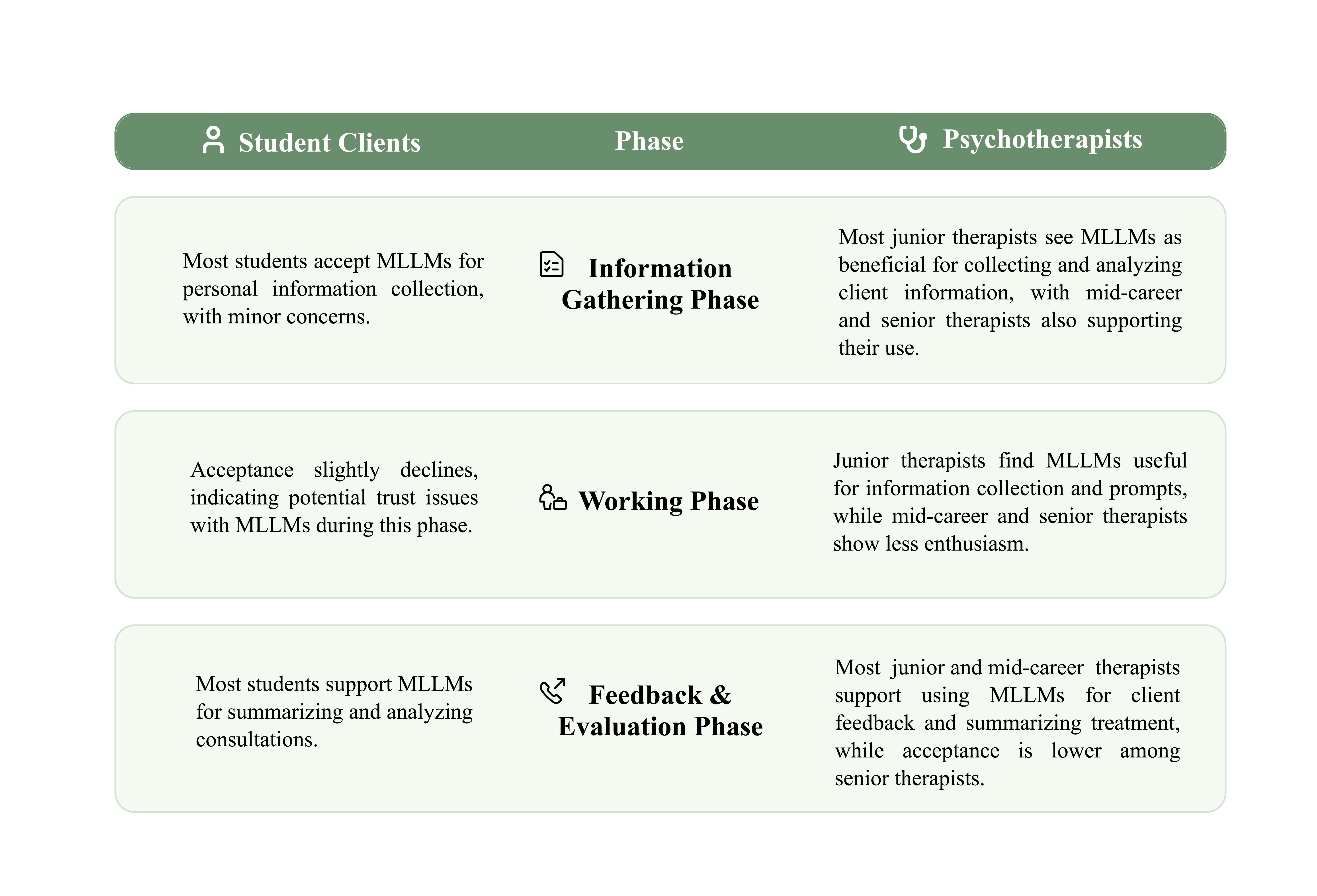}\\
\end{center}
\vspace{-4mm}
\caption{Summary of attitudes of student clients and psychotherapists on MLLM in different phases of therapy.}\label{f6}
\vspace{-4mm}
\end{figure*}

As summarized in Figure \ref{f6}, overall, all three studies indicated that both therapists and student users hold a positive attitude towards the application of MLLM in psychological counseling tasks, believing that MLLM can provide valuable assistance in certain aspects of psychotherapy. However, as the level of expertise increases, therapists' attitudes tend to become more cautious. This shift may be attributed to differences in age and accumulated experience \citep{doh2015patterns}. This is especially evident in areas involving emotional support and complex decision-making, where experts may be conservative about MLLM's potential to replace human therapists. The acceptance of MLLM in psychological counseling is also influenced by the type of task and the user's mental health status. In structured tasks such as information gathering and feedback analysis, the technical capabilities of MLLM are widely recognized. These tasks mainly involve data processing and analysis, making users more likely to accept MLLM assistance as they can clearly perceive its efficiency and accuracy \citep{jin2024llms}. 

In more complex tasks, such as the working phase, MLLM is not only required to process data but also to engage in real-time decision-making and offer suggestions. Thus, the influence of relative status remains significant. As the result of the survey, patients are more inclined to accept MLLM which is perceived as having a higher social status, as they expect it to provide valuable insights in complex situations. This implies that users no longer view MLLM as a mere tool but as a "companion" with a certain degree of autonomy and insight. This further enhances the acceptance of MLLM in more complex tasks. This phenomenon can be explained by the Computers Are Social Actors theory \citep{nass1994computers}. Whether in structured or complex tasks, MLLM with higher social status is more likely to be accepted due to the expectations conferred upon it.

However, in tasks involving deep emotional interaction and autonomous decision-making (e.g., the working phase and autonomous therapy agents), users' acceptance of MLLM significantly decreases, particularly among those in the mental illness group, who exhibit strong resistance. This may be because such users rely more heavily on the emotional support and interaction provided by human therapists, and they worry that MLLM may not be able to provide sufficient understanding and care in these areas \citep{thoits2011mechanisms}. Although MLLM performs well in data processing, its lack of emotional intelligence may exacerbate these users' anxiety, leading to resistance towards MLLM's autonomous decision-making capabilities. Additionally, the survey result showed that patients with a strong sense of social identity, due to their social status and self-perception, tend to embrace new technologies, especially when they experience low levels of anxiety. In such cases, their acceptance of MLLM increases significantly. However, when users' anxiety levels are high, even if they have a strong social identity, anxiety tends to dominate their attitudes. It can be explained by the stronger willingness of people with anxiety to control their health than others \citep{shapiro1996controlling}, which might lead to higher concerns about the effectiveness of MLLM-based therapy. Conversely, users with low anxiety tend to be more emotionally stable and can evaluate the advantages of MLLM more rationally. Therefore, the enhancement of social identity further increases their acceptance of MLLM. This aligns with the theory of emotion-driven behavior in psychology, which posits that strong negative emotions, such as anxiety, can offset the positive influence of social status or other favorable factors on behavior \citep{franke2015advancing}. Combining previous research \citep{lattie2020designing} for digital mental health tools, users' identified social factors and roles should be considered in future MLLM designed for mental health.

\subsection{Multi-side users' function demand}
Given the capability of current MLLMs, users from both two sides of the therapeutic alliance expressed consistent function needs in MLLM-assisted psychotherapy. Overall, referring to the results of Study I and II, most users (20/35) agreed that MLLM should assist the therapists in the information gathering stage to collect and analyze basic information about the client. During the working phase, except for real-time emotion detection, users also suggested emergency management (10/35), which means MLLMs are expected to detect any physiological or psychological abnormalities that may occur in advance, in order to provide timely emergency assistance. In the feedback and evaluation phase, users' needs for MLLM focused on collecting client feedback and helping therapists summarise the treatment process in order to consolidate the results. In short, these functions mainly require excellent content comprehension and generation capabilities of MLLM, which are already achieved by most recent MLLMs in other domains \citep{wang2024evaluating,petridis2023anglekindling}. 

Moreover, the impact of client heterogeneity \citep{zilcha2015one,kaiser2022heterogeneity} on treatment outcomes was highlighted by both therapists and students, which is also in line with previous research \citep{zhou2022application,molli2022effectiveness}. In addition to assisting therapists in customizing treatment plans for different clients based on their historical treatment records, intelligent triage was also emphasized by both the therapists and student clients (16/35). Therapists expect the MLLM to \textit{"provide better service on the issues they specialize in"}, and clients also have noted that \textit{"I am willing to talk to some therapists even we just meet; but I struggle to open up to others even if they encourage me."} Therefore, if future MLLMs can integrate information from both therapists and clients to facilitate two-way matching, the quality of the current psychotherapy services can be improved. 

In addition, aligned with the results from the survey study, we found that students were placing higher anthropomorphic demands on MLLMs that aim to independently conduct daily counseling or automated psychotherapy. This is reasonable, as students suffering from varying degrees of mental issues agreed that, the lack of physical entities and images was an important barrier to their difficulties in establishing a relationship with MLLM without the presence of the human therapist. In past conversational therapy, human therapists could provide emotional support by nonverbal communication, such as physically touching the patient \citep{bonitz2008use} or conveying their empathic feelings through facial expressions and gestures \citep{foley2010nonverbal}. However, the existing MLLM is not yet sufficient, particularly when there is no human therapist cooperating with it; in other words, users expect future MLLM to go beyond the output of one modality, natural language, in the provision of mental health services. Furthermore, in terms of physical image requirements of MLLMs, a notable number of student users indicated that if they could customize it with their preferred image (e.g. a cute anime cartoon character, or a real-life favorite person), it would be more warm and comforting than cold, metallic machinery. However, we also note that previous researchers have mentioned that MLLM should not use the same image as a real person to avoid ethical and copyright issues \citep{romele2022images}. Therefore, how to design user-satisfying and ethical images in the future is a direction worthy of further research.

\subsection{Users' concerns to current products}
In general, the highly mentioned concerns expressed by both two sides of the current human therapeutic alliance at the university (therapists and student patients) are about the current MLLM's ability to provide emotional support, and ensure privacy and ethical security. Actually, these points have been mentioned in previous studies targeting AI tools with weaker capability \citep{ebert2019digital,grodniewicz2023waiting}. However, faced with a more capable MLLM, therapists and patients offer new explanations for their concerns from their own perspectives. 

Firstly, For therapists, given establishing deep emotional relationships and effectively managing transference is critical in psychotherapy \citep{hoglend2014exploration}, most therapists continue to express doubts about the ability of existing MLLM to manage empathy independently. Despite the more positive attitude shown by junior therapists, the ability of existing MLLM in this area has not yet met even the expectations of junior therapists. In terms of relationship building, the experts' questioning centered mainly on the patient's ability to successfully establish a deep emotional relationship with an MLLM belonging to a non-human being. The students went further and gave their reflections from the other side. The students noted that although existing MLLMs can perceive information from other modalities such as speech and vision, and have stronger content comprehension compared to previous unimodal AIs \citep{wang2024evaluating}, they nevertheless believe that there is still space for improvement in terms of intelligence capabilities (e.g., too rigid language, possible misrepresentation, lack of depth in conversation). In particular, I1 pointed out that the lack of real-life experiences made it difficult to reach the heart and make their answers convincing. Therefore, in addition to further improving the generation and content comprehension abilities, future MLLM for mental health could consider introducing real human experiences into the generated content, including previous similar cases and celebrity examples, to increase the persuasiveness and empathy of the responses.

Moreover, users' concerns about privacy and ethical security are in line with previous research findings \citep{yoo2023discussing,tekin2023ethical,martinez2018ethical}. Specifically, compared to other scenarios (e.g. academic work \citep{wang2024young}), therapists usually need to encourage the client to overcome their stigma and shame, and try to break through their defense mechanism. Thus personal and private information will be inevitably included during the psychotherapy. In traditional therapeutic alliances, clients' privacy is ensured primarily through rigorous industry regularization and the training of therapists \citep{apaprivacy}. When MLLM is involved in existing therapeutic processes, one of the biggest challenges is ensuring privacy security and enabling users to trust its privacy protection mechanisms. In fact, a number of approaches have been proposed for securing data in AI models (e.g., privacy claim \citep{hamdoun2023ai}, federated learning \citep{khalil2024exploring}). To some extent, more effort has been invested in regulating AI than in regularizing human therapists (e.g., codes of action, legal texts, etc.). However, based on our findings, users remain skeptical about this. To explain this, we retrieved another interesting finding that student users raised concerns about MLLM's malicious manipulation of humans only when MLLM undertook daily conversational therapy and autonomous therapy. This somewhat reflects the fact that the majority of student users were more likely to believe in human groups, although two other interviewees expressed a higher level of trust in MLLM than humans. The positive attitudes of these two students may be related to their openness to the new technology, as well as to their experience of attending counseling (one clearly expressed a strong embrace of the new technology, the other from normal student group). Overall, in the college students group, who are overall more open to new technology \citep{broady2010comparison}, the majority of them still showed a relatively higher level of trust in humans. Therefore, we believe that more capable MLLMs in the future may need to first handle high resistance from human society if they aim to replace human therapists.

\subsection{Suggestions and Implications}

Based on our three studies, we conclude with several suggestions for future MLLM design for psychotherapy. First of all, the human therapy alliance is still currently in favor of considering MLLM as an adjunct to be involved in the psychotherapeutic process, and they have different demands at different phases of treatment. Especially in the phase of information gathering and feedback and evaluation, for the ease of function development, current developers can prioritize functions like information collection, storage, and analysis that serve therapists. As for the intelligent triage function applied in the information collection stage, although there is a lot of research work on the matching mechanism, considering the different resources of each service institution and the changes in the supply and demand of mental health services in the real world, it may be necessary to further improve the client-therapist matching scheme before the implementation of the functions. Further, while real-time emotion recognition for contingency situation management in the working phase was ignored by senior therapists, it is something that can still be considered in future product design as it is demanded by both younger therapists and student clients, especially when this is not too complicated from a technical point of view. However, it is worth noting that in the process of assisted therapy, due to the need to collect a large amount of privacy information, MLLM should more effectively highlight its efforts on privacy protection to all users.

For the MLMM products that can carry out therapy independently, daily conversational therapy services are relatively more accepted among users; while users are still skeptical of autonomous therapy agents. From an anthropomorphic perspective, the image of the MLLM, including a personalized and customized avatar and a physical image that can perform non-verbal emotional expression and soothing, is the core demand. The realization of avatars through the integration of virtual reality or digital human technologies with MLLM is a viable route. The implementation of physical images, on the other hand, still requires a great deal of effort in terms of the current level of technological development in order to achieve a capability close to that of a human therapist. Further, the accessibility can mainly be hindered by the high cost of MLLMs, especially when users hope to achieve anytime-anywhere access to MLLMs and to keep a low (especially lower than the cost of human therapists) expense at the same time. Other identified obstacles include human self-identity and the defense of self-interest (e.g., loss of job) to allow humans to fully trust and accept the AI agent represented by MLLM as a substitute for humans in providing mental health services. From the academic perspective, we call for the expansion of existing technology acceptance models (e.g., TAM \citep{davis1989technology}, UTAUT \citep{venkatesh2012consumer}) for psychotherapeutic scenarios to include socio-psychology factors (e.g., self-identity, relative status perception). 

In all, we believe that the easiest way to implement AI-assisted tools at this stage is to design AI-assisted tools and revolutionize the existing treatment process to improve the efficiency and effectiveness of the treatment, so that the traditional two-party human therapeutic alliance can be developed into a new three-party therapist-client-AI therapeutic alliance.

\section{Limitations}

Our study aimed to extract consensus from both sides of the human therapeutic alliance, but our sampling was only from China. We acknowledge the impact of individual, cultural, and socio-economic differences on our findings, so more evaluation work in different countries and cultural contexts is still needed. In addition, the state-of-the-art MLLM benchmarked in this paper is GPT-4o. Although we have tried to avoid limiting the interview process and discussion of the results on the current single model, users' views may still change as the language model improves. Finally, the research methodology of this paper is interviews and questionnaires. Empirical experiments are still needed to evaluate the existing treatment workflow with integrated AI. Future empirical studies could consider conducting experiments based on the scenarios/phases defined in this paper, then quantifying and giving different priorities to the different needs of users.

\section{Conclusions}
In this paper, we conducted three studies to understand the opportunities and challenges of integrating MLLM into campus psychotherapy, based on the feedback from both psychotherapists and student clients. Our finding shows that serving the MLLM as an assistant under the supervision of human therapists is still mostly preferred at this stage, which can make better use of MLLM's multimodal perception, strong content understanding, and generation capabilities to enhance the efficiency and effectiveness of the treatment. At the same time, users highly expect intelligent triage matching and real-time mental/physiological state recognition. Whereas when the MLLM completes therapy independently, whether it is daily counseling or full psychotherapy, users still distrust the MLLMs and have privacy and ethics concerns. Finally, we found that the driving factors of users' attitudes toward MLLM in mental health encompassed users' social identification (as human beings or not) and their perceived relative status of MLLM (servant or mentor). Future work could further expand the existing technology acceptance model from a psychosocial perspective, to facilitate the digital and intelligent campus mental health care.

\section*{Acknowledgement}
This work was supported by the Guangzhou Municipal Science and Technology Project (No. 2023A03J0011), the Guangzhou Science and Technology Program City-University Joint Funding Project (No. 2023A03J0001), and the General Project of Education of National Social Science Foundation (No. BIA220072).  The contribution of Jiyao Wang, Youyu Sheng, and Qihang He throughout the entire work is equal. 

\section*{Disclosure statement}
No potential conflict of interest was reported by the author(s).

\appendix

\section{Outline of the focus group interview}
\label{app:A}
\begin{enumerate}
    \item Have you heard of the large language model and what are your attitudes toward its use in counseling?
    \item What do you think are the general phases of a complete counseling session in college? Which of these phases can MLLM be used?
    \item What functions or features do you think MLLM needs to have when it is used for counseling?
    \item What do you think might be the problems in applying MLLM in psychotherapy?
\end{enumerate}

\section{Outline of the semi-structured interview}
\label{app:B}

\begin{enumerate}
  \item Please introduce yourself.
  \item Have you ever heard of the MLLM? Do you usually talk to MLLM when worries or problems arise in your mind? Why?
  \item If there is an MLLM that can assist human therapists in their counseling work, what is your attitude towards this? What functions or features do you think this AI needs to have? What are your concerns? (If you have not received counseling, please imagine based on your life experience).
  \begin{itemize}
      \item Before you start talking to the therapist, are you comfortable with the counselor using MLLM to collect and analyze your personal information in advance?
      \item During your conversations with the therapist, are you comfortable with the counselor using MLLM to observe you and provide advice to them?
      \item At the end of your conversation, are you comfortable with the therapist using MLLM to summarise and analyze the content and effectiveness of the counseling session?
  \end{itemize}
  \item If there is an MLLM who can act like a mentor or a friend in your life to solve your various mental problems and troubles, what is your attitude towards this? What functions or features do you think this MLLM needs to have? What issues would you worry about?
  \item If there is a powerful agent in the future who can do counseling work like a human therapist, what is your attitude towards this? What functions or features do you think this agent needs to have? What issues would you worry about?
\end{enumerate}

\bibliographystyle{elsarticle-harv} 
\bibliography{ref}

\end{document}